\documentclass[fleqn,twoside,11pt]{article}

\usepackage{amssymb}
\usepackage{graphicx}

\usepackage{amsmath}

\begin{document}
\noindent
{\sf University of Shizuoka}

\hspace*{13cm} {\large US-05-01}

\vspace{2mm}

\begin{center}

{\Large\bf Quark and Lepton Mass Matrix Structures}\\[.2in]
{\Large\bf Suggested by the Observed Unitary Triangle Shape}

\vspace{3mm}

{\bf Yoshio Koide}

{\it Department of Physics, University of Shizuoka, 
52-1 Yada, Shizuoka 422-8526, Japan\\
E-mail address: koide@u-shizuoka-ken.ac.jp}

\date{\today}
\end{center}

\vspace{3mm}
\begin{abstract}
Under the hypothesis that the $CP$ violating phase parameter
$\delta$ in the CKM matrix $V$ takes own value so that the radius
$R(\delta)$ of the circle circumscribed about the unitary
triangle takes its minimum value, possible phase conventions
of the CKM matrix are investigated.  We find that two of the
9 phase conventions can give favorable predictions for
the observed shape of the unitary triangle.  One of the 
successful two suggests phenomenologically interesting structures
of the quark and lepton mass matrices, which lead to 
$|V_{us}|\simeq \sqrt{m_d/m_s}=0.22$, $|V_{ub}|\simeq \sqrt{m_u/m_t}
=0.0036$ and $|V_{cb}|\simeq \sqrt{m_c/2m_t}=0.043$ for the CKM
matrix $V$, and to $\sin^2 2\theta_{atm}=1$, $\tan^2 \theta_{solar}
\simeq |m_{\nu 1}/m_{\nu 2}|$ and $|U_{13}|\simeq \sqrt{m_e/2 m_\tau}$
for the lepton mixing matrix $U$ under simple requirements for the 
textures.
\end{abstract}



\vspace{5mm}


\noindent{\large\bf 1 \ Introduction} \ 

Recent remarkable progress of the experimental $B$ physics
\cite{B} has put the shape of the unitary
triangle in the quark sector within our reach. 
The world average value of the angle $\beta$ \cite{PDG04} which has been
obtained from $B_d$ decays is
$$
\sin 2\beta = 0.736 \pm 0.049 \ \ 
\left( \beta = 23.7^\circ {}^{+2.2^\circ}_{-2.0^\circ} \right) ,
\eqno(1.1)
$$
and the best fit \cite{PDG04} for the Cabibbo-Kobayasi-Maskawa (CKM) 
matrix \cite{Cabibbo,KM} $V$ also gives 
$$
\gamma = 60^{\circ} \pm 14^{\circ} \ , \ \ \ 
\beta = 23.4^{\circ} \pm 2^{\circ} \ ,
\eqno(1.2)
$$
where the angles $\alpha$, $\beta$ and $\gamma$ are defined by
$$
\alpha = {\rm Arg} \left[-\frac{V_{31} V^{*}_{33}}{V_{11} V^{*}_{13}} \right]
 , \ \ \ 
\beta = {\rm Arg} \left[-\frac{V_{21} V^{*}_{23}}{V_{31} V^{*}_{33}} \right]
 , \ \ 
\gamma = {\rm Arg} \left[-\frac{V_{11} V^{*}_{13}}{V_{21} V^{*}_{23}} \right]
 .
\eqno(1.3)
$$
We are interested 
what logic can give the observed magnitude of the $CP$ violation. 

Usually, we assume a peculiar form of the quark mass matrices at the start, 
and thereby,
we predict a magnitude of the $CP$ violation and a shape of the
unitary triangle.
However, recently, the author \cite{Koide_maxCP} has investigated 
a quark mass matrix model on the basis of an inverse procedure: 
by noticing that predictions based on the maximal $CP$ violation hypothesis
\cite{maxCP}
depend on the phase convention, the author has, at the start, 
investigated what phase conventions can give favorable predictions 
of the unitary triangle, and then he has investigated what quark mass
matrices can give such a phase convention of the CKM matrix.
Here, we have assumed that the three rotation angles in the CKM
matrix $V$ are fixed by the observed values of $|V_{us}|$, $|V_{cb}|$
and $|V_{ub}|$, and only the $CP$ violating phase parameter $\delta$
is free.

There are, in general, 9 cases \cite{FX-9V} for the phase convention 
of the CKM matrix.
When we define the expression of the CKM matrix $V$ as
$$
V = V(i,k) \equiv R^T_i P_j R_j R_k \ \ \ \ \ (i \neq j \neq k) ,
\eqno(1.4)
$$
where 
$$
R_1 (\theta) = \left(
\begin{array}{ccc}
1 & 0 & 0 \\
0 & c & s \\
0 & -s & c 
\end{array} \right) , \ \ \ \ 
R_2 (\theta) = \left(
\begin{array}{ccc}
c & 0 & s \\
0 & 1 & 0 \\
-s & 0 & c 
\end{array} \right) , \ \ \ \ 
R_3 (\theta) = \left(
\begin{array}{ccc}
c & s & 0 \\
-s & c & 0 \\
0 & 0 & 1 
\end{array} \right) ,
\eqno(1.5)
$$
($s=\sin\theta$ and $c=\cos\theta$)
and $P_1 = {\rm diag} (e^{i \delta}, \ 1, \ 1)$, 
$P_2 = {\rm diag} (1, \ e^{i \delta}, \ 1)$, and 
$P_3 = {\rm diag} (1, \ 1, \ e^{i \delta})$, 
then the rephasing invariant quantity \cite{J} $J$  is given by
$$
J=\frac{|V_{i1}||V_{i2}||V_{i3}||V_{1k}||V_{2k}||V_{3k}|}
{(1 -|V_{ik}|^2 ) |V_{ik}|} \sin \delta \ .
\eqno(1.6)
$$
And also, angles $\phi_\ell$ ($\ell=1,2,3$;
$\phi_1=\beta$, $\phi_2=\alpha$ and $\phi_3=\gamma$ in the
conventional angle notations) of the triangle $\bigtriangleup^{(31)}$
for the unitary condition
$$
V^{*}_{ud} V_{ub} + V^{*}_{cd} V_{cb} + V^{*}_{td} V_{tb} =0.
\eqno(1.7)
$$
are given by the formula
$$
\sin\phi_\ell=\frac{|V_{i1}||V_{i2}||V_{i3}|
|V_{1k}||V_{2k}||V_{3k}|\sin\delta}{|V_{m1}||V_{m3}|
|V_{n1}||V_{n3}|(1 -|V_{ik}|^2 ) |V_{ik}|} ,
\eqno(1.8)
$$
where $(\ell, m, n)$ is a cyclic permutation of (1,2,3).
(Note that the 5 quantities $|V_{i1}|$, $|V_{i2}|$, $|V_{i3}|$,
$|V_{1k}|$, $|V_{2k}|$ and $|V_{3k}|$ in the expression
$V(i,k)$ are independent of the phase parameter $\delta$.
In other words, only the remaining 4 quantities are dependent 
of $\delta$.)
The author \cite{Koide_maxCP} has found that phase conventions 
which lead to 
successful predictions under the maximal $CP$ violation hypothesis 
\cite{maxCP} are only two: the original Kobayasi-Maskawa phase 
convention \cite{KM} $V(1,1)$ and the Fritzsch-Xing phase 
convention \cite{V33}.  

The author \cite{Koide_maxCP} has also pointed out that the phenomenological
success of the expression $V(3,3)$ suggests a quark mass 
matrix form \cite{V33,Xing03} 
$$
M_q = P_1^q R_1^q R_3^q D_q R_3^{qT} R_1^{qT} P_1^{q\dagger}
\ \ \ (q=u,d),
\eqno(1.9)
$$
where 
$$
D_q = {\rm diag}(m_{q1}, m_{q2}, m_{q3}).
\eqno(1.10)
$$
The quark mass matrix form (1.9) leads to the well-known
successful prediction \cite{Vus} 
$$
|V_{us}|\simeq \sqrt{m_d/m_s}
\eqno(1.11)
$$
under the texture-zero requirement $(M_d)_{11}=0$, while
the texture-zero requirement $(M_u)_{11}=0$
predicts ${|V_{ub}|}/{|V_{cb}|} \simeq \sqrt{{m_u}/{m_c}}
\simeq 0.059$ (we have used values \cite{q-mass} at $\mu =m_Z$ 
as quark mass values), which is
in poor agreement with the observed value
\cite{PDG04} ${|V_{ub}|}/{|V_{cb}|} = 0.089^{+0.015}_{-0.014}$.

Therefore, in the present paper, we will investigate another
possibility instead of the maximal $CP$ 
violation hypothesis.
In Sec.~2, by assuming that the $CP$ violating phase parameter
$\delta$ in the CKM matrix $V$ takes own value so that the radius
$R(\delta)$ of the circle circumscribed about the unitary
triangle takes its minimum value, we will find that only two types 
$V(2,3)$ and $V(2,1)$ can give favorable predictions for
the observed shape of the unitary triangle.
 Stimulated by this result, in Secs.~3 and 4, we will assume that 
the quark mixing matrix $V=U_u^\dagger U_d$ and lepton mixing matrix 
$U=U_e^\dagger U_\nu$ 
are given by the type $V(2,3)$, and we will obtain successful predictions
$|V_{us}|\simeq \sqrt{m_d/m_s}=0.22$, $|V_{ub}|\simeq \sqrt{m_u/m_t}
=0.0036$ and $|V_{cb}|\simeq \sqrt{m_c/2m_t}=0.043$ for the CKM
matrix $V$, and $\sin^2 2\theta_{23}=1$, $\tan^2 \theta_{12}
\simeq |m_{\nu 1}/m_{\nu 2}|$ and $|U_{13}|\simeq \sqrt{m_e/2 m_\tau}$
for the lepton mixing matrix \cite{MNS} $U$ under simple requirements 
for mass matrix textures.
Finally, Sec.~5 will be devoted to concluding remarks.

\vspace{2mm}

\noindent{\large\bf 2 \ Ansatz for the unitary triangle} \ 

Of the three unitary triangles $\bigtriangleup^{(ij)}$
[$(ij)=(12)$, $(23)$, $(31)$] which denote the unitary conditions
$$
\sum_k V^{*}_{ki} V_{kj} = \delta_{ij} \ ,
\eqno(2.1)
$$
we usually discuss the triangle $\bigtriangleup^{(31)}$, i.e.
$$
V^{*}_{ud} V_{ub} + V^{*}_{cd} V_{cb} + V^{*}_{td} V_{tb} =0,
\eqno(2.2)
$$
because the triangle $\bigtriangleup^{(31)}$ is the most useful 
one for the experimental studies.
Seeing from the geometrical point of view, the triangle 
$\bigtriangleup^{(31)}$ has the plumpest shape compared with
other triangles $\bigtriangleup^{(12)}$ and $\bigtriangleup^{(23)}$, 
so that the triangle $\bigtriangleup^{(31)}$  has the shortest radius 
$(R_{(31)})_{mini}$ of the circumscribed circle compared with 
the other cases $\bigtriangleup^{(12)}$ and $\bigtriangleup^{(12)}$.

Therefore, let us put the following assumption:
the phase parameter $\delta$ takes the value so that the circumscribed
circle $R_{(31)}(\delta)$ takes its minimum value.

The radius $R_{(31)}(\delta)$  is given by the sine rule
$$
\frac{r_1}{\sin\phi_1}=\frac{r_2}{\sin\phi_2}=
\frac{r_3}{\sin\phi_3}= 2 R_{(31)},
\eqno(2.3)
$$
where
$$
r_1 =|V_{13}| |V_{11}|, \ \ r_2 =|V_{23}| |V_{21}|, \ \ 
r_3 =|V_{33}| |V_{31}|,
\eqno(2.4)
$$
and the angles $(\phi_1, \phi_2, \phi_3)\equiv (\beta, \alpha, \gamma)$ 
are defined by Eq.~(1.3).  
Note that in the expression $V(i,k)$ the side $r_i$ is independent of
the parameter $\delta$. 
Therefore, the minimum of the radius $R_{(31)}(\delta)$ means 
the maximum of $\sin\phi_i(\delta)$.
We put a further assumption:
the phase parameter $\delta$ takes own value so that $\sin\phi_i(\delta)$
takes its maximal value, i.e. $\sin\phi_i=1$.

Then, we find that 6 cases of the 9 cases $V(i,j)$  except for 
$V(3,3)$, $V(2,3)$ and $V(2,1)$ cannot give  $\sin\phi_i=1$
(i.e. $|\sin\phi_i| < 1$) 
under the observed values  \cite{PDG04} of $|V_{us}|$, $|V_{cb}|$ 
and $|V_{ub}|$
$$
|V_{us}| = 0.2200 \pm 0.0026 , \ \
|V_{cb}| = 0.0413 \pm 0.0015 , \ \
|V_{ub}| = 0.00367 \pm 0.00047 .
\eqno(2.5)
$$
Of the three candidates, the case $V(3,3)$ ruled out, because
the requirement $\sin\phi_3=1$ (i.e. $\gamma/2 =\pi$) 
disagrees with the observed value (1.2).
Therefore, the possible candidates for the requirement
$\sin\phi_i=1$ are only $V(2,3)$ and $V(2,1)$.
The requirement $\sin\phi_i =1$  means
the requirement $\sin\alpha =1$ in $V(2,3)$ and $V(2,1)$.
{}From the relation (2.3), we obtain
$$
\sin\beta = \frac{r_1}{r_2} \sin\alpha =\frac{|V_{13}||V_{11}|}{
|V_{23}||V_{21}|} , \ \ 
\sin\gamma = \frac{r_3}{r_2} \sin\alpha =\frac{|V_{33}||V_{31}|}{
|V_{23}||V_{21}|} .
\eqno(2.6)
$$
For example, the case $V(2,3)$  predicts
$$
|V_{td}|= \left(8.36 ^{+0.24}_{-0.27}\right) \times 10^{-3},  \ \  
\beta = 23.2^{\circ} \pm 0.1^{\circ} ,  \ \ 
\gamma = {66.8^{\circ}}^{+3.8^\circ}_{-4.3^\circ} , 
\eqno(2.7)
$$
with $\delta ={113.2^\circ}^{-3.8^\circ}_{+4.3^\circ}$
from the requirement $\alpha= 90^\circ$  and the observed values (2.5).
The predictions (2.7) are in good agreement with the observed values 
(1.1) and (1.2).
For the case $V(2,1)$, we obtain the same numerical results (2.7)
(but with a different value of $\delta$).
As far as the phenomenology of the unitary triangle is concerned,

we cannot determine which case is favor.

However, as we see in the next section, the case $V(2,3)$ suggests
an interesting texture of the quark mass matrices, which leads to
predictions $|V_{us}| \simeq \sqrt{m_d/m_s}$,
$|V_{ub}| \simeq \sqrt{m_u/m_t}$ and $|V_{cb}| \simeq \sqrt{m_c/2m_t}$
under a simple ansatz.
For the case $V(2,1)$, we cannot obtain such an interesting texture.

\vspace{2mm}

\noindent{\large\bf 3 \  Speculation on the quark mass matrix form} \ 

As we have assumed in the previous paper, the successful expression
$V(i,k)$ suggests the following situation:
The phase factors in the quark mass matrices $M_f$ \ $(f=u,d)$ are 
factorized by the phase matrices $P_f$ as
$$
M_f = P^{\dagger}_{fL} \widetilde{M}_f P_{fR} \ ,
\eqno(3.1)
$$
where $P_f$ are phase matrices and $\widetilde{M}_f$ are real matrices. The
real matrices $\widetilde{M}_f$ are diagonalized by rotation (orthogonal)
matrices $R_f$ as
$$
R^{\dagger}_f \widetilde{M}_f R_f = D_f 
\equiv {\rm diag} (m_{f1}, \ m_{f2}, \ m_{f3} ),
\eqno(3.2)
$$
[for simplicity, we have assumed that $M_f$ are Hermitian (or symmetric) 
matrix, i.e. $P_{fR} = P_{fL}$ (or $P_{fR} = P_{fL}^{\ast}$)], so that 
the CKM matrix $V$ is given by
$$
V = R^T_u P R_d \ ,
\eqno(3.3)
$$
where $P = P^{\dagger}_{uL} P_{dL}$. 
The quark masses $m_{fi}$ are only 
determined by $\widetilde{M}_f$. 
In other words, the rotation parameters are
given only in terms of the quark mass ratios, and independent 
of the $CP$ violating phases.
In such a scenario, the $CP$ violation parameter 
$\delta$ can be adjusted  without changing the quark mass values.

For example, the case $V(2,3)$ suggests the quark mass matrix structures
$$
\begin{array}{c}
\widetilde{M}_u = R_1(\theta^u_{23}) R_2(\theta_{13}^u) 
D_u R^T_2(\theta_{13}^u)
R_1^T (\theta^u_{23}) \ , \\
\widetilde{M}_d = R_1(\theta^d_{23}) R_3(\theta_{12}^d) 
D_d R^T_3(\theta_{12}^d)

R_1^T (\theta^d_{23}) \ ,
\end{array}
\eqno(3.4)
$$
with $\theta_{23}=\theta^d_{23}-\theta^u_{23}$. 
The explicit forms of $V(2,3)$, $\widetilde{M}_u$ and 
$\widetilde{M}_f$ are given as follows:
$$
V(2,3) = R_2^T (\theta_{13}^u) P_1 (\delta) R_1 (\theta_{23}) 
R_3(\theta^d_{12})
$$
$$
= \left(
\begin{array}{ccc}
e^{i \delta} c^u_{13} c^d_{12} - s_{23}s^u_{13} s^d_{12} 
& e^{i \delta} c^u_{13} s^d_{12} + s_{23} s^u_{13} c^d_{12} &
 -c_{23} s^u_{13}  \\
-c_{23} s^d_{12} & c_{23} c^d_{12} &  s_{23} \\
e^{i \delta} s^u_{13} c^d_{12} + s_{23} c^u_{13} s^d_{12} &
e^{i \delta} s^u_{13} s^d_{12} - s_{23} c^u_{13} c^d_{12} &
c_{23} c^u_{13} \\
\end{array}\right) ,
\eqno(3.5)
$$

$$
\widetilde{M}_u = \left(
\begin{array}{cc}
m_{u1} (c^{u}_{13})^2 + m_{u3} (s_{13}^{u})^2 &
(m_{u3} - m_{u1}) c^u_{13} s^u_{13} s^u_{23}  \\
(m_{u3} - m_{u1}) c^u_{13} s^u_{13} s^u_{23} &
[ m_{u1} (s^{u}_{13})^2 + m_{u3} (c^{u}_{13})^2] (s_{23}^{u})^2 
+ m_{u2} (c^{u}_{23})^2  \\
(m_{u3} - m_{u1}) c^u_{13} s^u_{13} c^u_{23} &
[  m_{u1} (s_{13}^{u})^2 + m_{u3} (c_{13}^{u})^2-m_{u2} ] c^u_{23} s^u_{23} 
\end{array} \right.
$$
$$
\left.
\begin{array}{c}
(m_{u3} - m_{u1}) c^u_{13} s^u_{13} c^u_{23} \\ 
\left[  m_{u1} (s_{13}^{u})^2 + m_{u3} (c_{13}^{u})^2-m_{u2}  \right] 
c^u_{23} s^u_{23} \\ 
\left[ m_{u1} (s^{u}_{13})^2 + m_{u3} (c^{u}_{13})^2 \right] (c_{23}^{u})^2 
+ m_{u2} (s^{u}_{23})^2
\end{array} \right) ,
\eqno(3.6)
$$

$$
\widetilde{M}_d = \left(
\begin{array}{cc}
m_{d1} (c^{d}_{12})^2 + m_{d2} (s_{12}^{d})^2 &
(m_{d2} - m_{d1}) c^d_{12} s^d_{12} c^d_{23}   \\
(m_{d2} - m_{d1}) c^d_{12} s^d_{12} c^d_{23} &
[ m_{d1} (s^{d}_{12})^2 + m_{d2} (c^{d}_{12})^2] (c_{23}^{d})^2 
+ m_{d3} (s^{d}_{23})^2  \\
-(m_{d2} - m_{d1}) c^d_{12} s^d_{12} s^d_{23} &
[ m_{d3} - m_{d2} (c_{12}^{d})^2 - m_{d1} (s_{12}^{d})^2] c^d_{23} s^d_{23} 
\end{array} \right.
$$
$$
\left.
\begin{array}{c}
-(m_{d2} - m_{d1}) c^d_{12} s^d_{12} s^d_{23}  \\ 
\left[ m_{d3} - m_{d2} (c_{12}^{d})^2 - m_{d1} (s_{12}^{d})^2\right] 
c^d_{23} s^d_{23} \\ 
\left[ m_{d1} (s^{d}_{12})^2 + m_{d2} (c^{d}_{12})^2\right] (s_{23}^{d})^2 
+ m_{d3} (c^{d}_{23})^2
\end{array} \right) .
\eqno(3.7)
$$
In the mass matrix (3.7), the ansatz $(\widetilde{M}_d)_{11} =0$ leads to 
the well--known relation (1.11).
On the other hand, for the up-quark mass matrix (3.6), 
the constraint $(\widetilde{M}_u)_{11} =0$ leads to the relation
$$
|V_{ub}| \simeq s_{13}^u \simeq \sqrt{\frac{m_u}{m_t}} =0.0036,
\eqno(3.8)
$$
which is in excellent agreement with the experimental value (2.5).
For the case $V(2,1)$, we cannot obtain such a simple relation.
Therefore, we will concentrate further investigation on the $V(2,3)$ 
model with the  quark mass matrices (3.6) and (3.7).

In order to fix the value of $\theta_{23}$, we put an ansatz
$$
\frac{(\widetilde{M}_u)_{23}}{(\widetilde{M}_u)_{22}}
= -\frac{(\widetilde{M}_u)_{13}}{(\widetilde{M}_u)_{12}}.
\eqno(3.9)
$$
At present, there is no theoretical reason for the constraint
(3.9).  It is pure phenomenological ansatz.
The requirement (3.9) leads to
$$
s_{23}^u = \sqrt{\frac{ -m_{u2}/2}{m_{u3}+m_{u1}-m_{u2}}}
\simeq \sqrt{\frac{m_c}{2 m_t}} = 0.043 ,
\eqno(3.10)
$$
which is in good agreement of the observed value of $|V_{cb}|$

in (2.5).
If we assume a constraint 
$$
(\widetilde{M}_d)_{23} = (\widetilde{M}_d)_{13} = 0 ,
\eqno(3.11)
$$
which corresponds to a special case in a requirement analogous 
to (3.9), we obtain $s_{23}^d=0$, so that we can 
obtain a successful prediction
$$
|V_{cb}| = s_{23} =|s_{23}^u| \simeq \sqrt{\frac{m_c}{2 m_t}} = 0.043 .
\eqno(3.12)
$$

Although, at present, the origins of the constraints (3.9) and
(3.11) are unknown, in the next section, we will find that similar
requirements for the lepton sector also lead to successful
predictions.

\vspace{2mm}

\noindent{\large\bf 4 \ Application to the lepton sector} \ 

If we suppose the correspondence $M_u \leftrightarrow M_\nu$
and $M_d \leftrightarrow M_e$ for the lepton mass matrices
$(M_\nu, M_e)$, the lepton mixing matrix \cite{MNS} $U=U_e^\dagger U_\nu$ 
will be given by the type $V(3,2)$.  However, the case gives
a wrong prediction $|U_{12}|= |c_{23} s^e_{12}| < \sqrt{m_e/m_\mu}$
under the constraint $(M_e)_{11}=0$.
Although it does not need to adhere the constraint $(M_e)_{11}=0$,
phenomenologically, it is more interesting to assume that the
lepton mixing matrix $U$ is also described by the type $V(2,3)$,
and not by the type $V(3,2)$.

In the case $U=V(2,3)$, correspondingly to the expression
$$
U = R_2^T (\theta_{13}^e) P_1 (\delta) R_1 (\theta_{23}) 
R_3(\theta_{12}^\nu),
\eqno(4.1)
$$
the lepton mass matrices are given by the structures
$$
\begin{array}{c}
\widetilde{M}_e = R_1(\theta^e_{23}) R_2(\theta_{13}^e) D_e 
R^T_2(\theta_{13}^e)
R_1^T (\theta^e_{23}) \ , \\
\widetilde{M}_\nu = R_1(\theta^\nu_{23}) R_3(\theta_{12}^\nu) 
D_\nu R^T_3(\theta_{12}^\nu)
R_1^T (\theta^\nu_{23}) \ ,
\end{array}
\eqno(4.2)
$$
with $\theta_{23}=\theta^\nu_{23}-\theta^e_{23}$. 

Similarly to the quark sector, we put the following
constraints: 
$$
(M_e)_{11}=0 ,
\eqno(4.3)
$$ 
$$
(M_\nu)_{11} = 0 ,
\eqno(4.4)
$$
$$
\frac{(\widetilde{M}_\nu)_{23}}{(\widetilde{M}_\nu)_{22}}
= -\frac{(\widetilde{M}_\nu)_{13}}{(\widetilde{M}_\nu)_{12}},
\eqno(4.5)
$$
$$
(\widetilde{M}_e)_{23} = (\widetilde{M}_e)_{13} = 0 .
\eqno(4.6)
$$
The requirements (4.3) and (4.4) lead to familiar relations
$$
\frac{s^e_{13}}{c^e_{13}} = \sqrt{\frac{m_e}{m_\tau}}
=0.0167 , \ \ \ 
\frac{s^\nu_{12}}{c^\nu_{12}} = \sqrt{\frac{-m_{\nu 1}}{m_{\nu 2}}},
\eqno(4.7)
$$
respectively.
The requirement (4.6) also leads to a similar result
$$
c^e_{23}=0 \ \ \left( \theta_{23}^e=\frac{\pi}{2} 
\right),
\eqno(4.8)
$$
(but note that the result is different from the result $s_{23}^d=0$ 
in the down-quark mass matrix).
On the other hand, correspondingly to Eq.~(3.10), the requirement (4.5) 
lead to a relation
$$
c_{23}^\nu = \sqrt{\frac{ m_{\nu 3}/2}{m_{\nu 3} -m_{\nu 2}-m_{\nu 1}}}.
\eqno(4.9)
$$
If we suppose $m_{\nu 3}^2 \gg m_{\nu 2}^2 > m_{\nu 1}^2$,
we obtain
$$
\theta_{23}^\nu = \frac{\pi}{4} -\varepsilon  ,
\eqno(4.10)
$$
where
$$
\varepsilon \simeq  \frac{m_{\nu 2} + m_{\nu 1}}{2 m_{\nu 3}} .
\eqno(4.11)
$$
By using the observed  values $\Delta m^2_{32} \simeq 2.8 \times 10^{-3}$
eV$^2$ \cite{K2K04}, $\Delta m^2 _{21} = (7.1^{+1.2}_{-0.6}) 
\times 10^{-5}$ eV$^2$ and $\theta_{solar}=32.5^{+2.4}_{-2.3}$ degrees 
\cite{SNO04}, we estimate
$$
\begin{array}{l}
m_{\nu 3} \simeq \sqrt{\Delta m^2_{32}} = 0.053 \ {\rm eV}, \\ 
m_{\nu 2} =\sqrt{ \Delta m^2_{21}/(1-\tan^2 \theta_{21})} = 0.011
\ {\rm eV}, \\ 
-m_{\nu 1} =  m_{\nu 2} \sqrt{\tan^2 \theta_{21}} = 0.0069
\ {\rm eV}.
\end{array}
\eqno(4.12)
$$
Therefore, the deviation $\varepsilon$ from $\theta_{23}^\nu=\pi/4$
is
$$
\varepsilon = \frac{m_{\nu 2} - |m_{\nu 1}|}{2 m_{\nu 3}} =
0.038 \ \ ( 2.2^\circ) ,
\eqno(4.13)
$$
so that the deviation $\varepsilon$ does not  visibly affect 
the prediction $\sin^2 2\theta_{23}=1$.
That is, the relation (4.9) naturally leads to
the prediction $\sin^2 2\theta_{atm} =1$.
Also the present model gives the prediction
$$
|U_{13}| = c_{23} s_{13}^e \simeq \sqrt{ \frac{m_e}{2 m_\tau} } = 0.012.
\eqno(4.14)
$$

\vspace{2mm}

\noindent{\large\bf 5 \ \ Concluding remarks} \ 

By assuming that three rotation angles 
in the CKM matrix $V$ are fixed by the observed values of
$|V_{us}|$, $|V_{cb}|$ and $|V_{ub}|$ and only a $CP$
violating phase parameter $\delta$ is free, and by putting
an ansatz that the phase parameter $\delta$ takes own value
so that the radius $R_{(31)}(\delta)$ of the circumscribed circle 
about the unitary triangle takes its minimal value, we have 
found that only the phase conventions $V(2,3)$ and $V(2,1)$ can 
predict the observed shape of the unitary triangle.

Of the two successful phase conventions,
we have noticed the case $V(2,3)$, which suggests the quark mass matrix 
structures
$$
\begin{array}{c}
\widetilde{M}_u = R_1(\theta^u_{23}) R_2(\theta_{13}^u) 
D_u R^T_2(\theta_{13}^u)
R_1^T (\theta^u_{23}) \ , \\
\widetilde{M}_d = R_1(\theta^d_{23}) R_3(\theta_{12}^d) 
D_d R^T_3(\theta_{12}^d)
R_1^T (\theta^d_{23}) \ ,
\end{array}
\eqno(5.1)
$$
where $\widetilde{M}_f$ is defined by Eq.~(3.1).
Under the phenomenological constraints, $(M_f)_{11}=0$ and 
Eqs.~(3.9) and  (3.11), we have obtained successful results 
$|V_{us}|\simeq \sqrt{m_d/m_s}=0.22$ [Eq.~(1.11)], 
$|V_{ub}|\simeq \sqrt{m_u/m_t}=0.0036$ [Eq.~(3.8)] and 
$|V_{cb}|\simeq \sqrt{m_c/2m_t}=0.043$ [Eq.~(3.12)].

Since we have assumed that the lepton mixing matrix $U=U_e^\dagger U_\nu$
is also given by the type $V(2,3)$, we have speculated that the lepton mass 
matrix structures are given by
$$
\begin{array}{c}
\widetilde{M}_e = R_1(\theta^e_{23}) R_2(\theta_{13}^e) D_e 
R^T_2(\theta_{13}^e)
R_1^T (\theta^e_{23}) \ , \\
\widetilde{M}_\nu = R_1(\theta^\nu_{23}) R_3(\theta_{12}^\nu) 
D_d R^T_3(\theta_{12}^\nu)
R_1^T (\theta^\nu_{23}) \ ,
\end{array}
\eqno(5.2)
$$
and we have naturally derived the relation $\sin^2 2\theta_{atm}=1$
under the requirements (4.3) -- (4.6) similar to the quark sector
and the neutrino mass hierarchy $m_{\nu 3}^2 \gg m_{\nu 2}^2 >
m_{\nu 1}^2$.
Note that the result $\theta_{23} \simeq \pi/2$ can be obtained
only for the case of the normal neutrino mass hierarchy, and it
can never be done for the inverse hierarchy, as seen in Eq.~(4.9).

By the way, the structures (5.1) and (5.2) ostensibly look like 
the quark-to-lepton correspondence 
$(M_u, M_d) \leftrightarrow (M_e, M_\nu)$.
However, we have put the constraints
$$
\frac{(\widetilde{M}_f)_{23}}{(\widetilde{M}_f)_{22}}
= -\frac{(\widetilde{M}_f)_{13}}{(\widetilde{M}_f)_{12}},
\eqno(5.3)
$$
on $M_u$ and $M_\nu$, and
$$
(\widetilde{M}_f)_{23} = (\widetilde{M}_f)_{13} = 0 ,
\eqno(5.4)
$$
on $M_d$ and $M_e$, respectively.
As the results of these constraints together with the constraint
$(M_f)_{11}=0$, we obtain the final forms of the quark and lepton mass 
matrices
$$
\widetilde{M}_f = \left(
\begin{array}{ccc}
0 & a & a \lambda \\
a & -b &  b \lambda \\
a \lambda & b \lambda & b(\lambda^2 -2)
\end{array} \right) ,
\eqno(5.5)
$$
for $M_u$ and $M_\nu$, and
$$
\widetilde{M}_f = \left(
\begin{array}{ccc}
0 & \sqrt{-m_{f1} m_{f2}} & 0 \\
\sqrt{-m_{f1} m_{f2}} & m_{f2} + m_{f1} & 0 \\
0 & 0 & m_{f3}
\end{array} \right) ,
\eqno(5.6)
$$
for $M_d$ and $M_e$, respectively.
Here, the expression (5.5) is taken as
$\lambda=c_{23}/s_{23} >1$ and $b>0$
for $M_u$, and as $-\lambda=s_{23}/c_{23} <1$ 
and $b<0$ for $M_\nu$. 
The expression (5.6) is taken as
$(m_{f1}, m_{f2}, m_{f3})=(-m_d, m_s, m_b)$ for $M_d$
and $(m_{f1}, m_{f2}, m_{f3})=(-m_e, m_\tau, m_\mu)$ 
for $M_e$.
Thus, in the final expressions (5.5) and (5.6),
the quark-to-lepton correspondence 
$(M_u, M_d) \leftrightarrow (M_\nu, M_e)$ is recovered.

If we attach great importance to the 
$(M_u, M_d) \leftrightarrow (M_\nu, M_e)$
correspondence, we may take  
$(m_{e1}, m_{e2}, m_{e3})=(-m_e, m_\mu, m_\tau)$
instead of $(m_{e1}, m_{e2}, m_{e3})=(-m_e, m_\tau, m_\mu)$
in the charged lepton mass matrix (5.6).
Then, the rotation matrix $R_1(\theta^e_{23})$ with
$\theta^e_{12}=\pi/2$ is replaced with the unit matrix
$R_1(0)$.  For this case, the prediction $\sin^2 2\theta_{23}=1$
is still unchanged, but the prediction (4.14) will be replaced
with
$$
|U_{13}| = c_{23} s_{13}^e \simeq \sqrt{ \frac{m_e}{2 m_\mu} } = 0.049.
\eqno(5.7)
$$

So far, we have not discussed a renormalization group equation
(RGE) effects on the mass matrices.
Since we know that the mass ratios $m_d/m_s$ and $m_u/m_c$ are
insensitive to the RGE effects, the textures $\widehat{M}_d$ and
$\widehat{M}_e$ given by Eq.~(5.6) are almost unchanged under
the RGE effects.
On the other hand, although the mass ratios $m_c/m_t$ and 
$m_s/m_b$ are, in general, sensitive to the RGE effects,
in the texture (5.5), the effects can be almost absorbed
into the factor $\lambda$ for the case $|y_t|^2 \gg |y_b|^2$.
Therefore, the texture (5.5) is also almost insensitive to the
RGE effects.

In conclusion, suggested by the observed shape of the unitary
triangle, we have found a unified structures of the quark and
lepton mass matrices, (5.5) and (5.6).
In other words, if we assume the structures (5.5) and (5.6)
for the quark and lepton mass matrices, we can obtain the
successful relations (1.11), (3.8), (3.12), (4.7), (4.10)
and (4.14) between the fermion mixing matrices and fermion
mass ratios.  Moreover, if we put the minimal $R_{(31)}(\delta)$
hypothesis, we can obtain the successful predictions (2.7)
for the shape of the unitary triangle.
However, at present, it is an open question what symmetries
can explain the forms (5.5) and (5.6).
And, the meaning of the minimal radius hypothesis is also
an open question.

\vspace{7mm}

\centerline{\large\bf Acknowledgments}

The author thanks Zhi-zhong Xing for his enjoyable
and helpful conversations.
This work was supported by the Grant-in-Aid for
Scientific Research, the Ministry of Education,
Science and Culture, Japan (Grant Number 15540283).


\end{document}